# RRDVCR: Real-Time Reliable Data Delivery Based on Virtual Coordinating Routing for Wireless Sensor Networks

Venkatesh, C S Sengar, K R Venugopal,
Department of Computer Science and Engineering,
University Visvesvaraya College of Engineering,
Bangalore University, Bangalore-560 001, India
venkateshm.uvce@bub.ernet.in

S S Iyengar
Florida International University, Miami, Florida, USA
L M Patnaik
Adjunct Professor and INSA Senior Scientist, National
Institute of Advanced Studies, Bangalore-12, India

*Abstract—* **Real-time industrial application requires routing protocol that guarantees data delivery with reliable, efficient and low end-to-end delay. Existing Routing(THVR) [13] is based velocity of Two-Hop Velocity and protocol relates two-hop velocity to delay to select the next forwarding node, that has overhead of exchanging control packets, and depleting the available energy in nodes. We propose a Real-Time Reliable Data delivery based on Virtual Coordinates Routing (RRDVCR) algorithm, based on the number of hops to the destination rather than geographic distance. Selection of forwarding node is based on packet progress offered by two-hops, link quality and available energy at the forwarding nodes. All these metric are co-related by dynamic co-relation factor. The proposed protocol uses selective acknowledgment scheme that results in lower overhead and energy consumption. Simulation results shows that there is about 22% and 9.5% decrease in energy consumption compared to SPEED [8] and THVR [13] respectively, 16% and 38% increase in packet delivery compared to THVR [13] and SPEED[8] respectively, and overhead is reduced by 50%.**

*Keywords- Link Reliability, Maximum Transmission Count [MTX], Virtual Coordinating Routing, Dynamic Co-relation Factor: f(rt).*

## I. Introduction (Heading 1)

Wireless Sensor Networks has a wide range application namely, intruder tracking, medical care, health diagnosis and fire monitoring [15]. Industrial communication community ap- plications demands rigid end-to-end delivery and reliability [10][13] under the constraints of wireless communication. QoS guarantees for Real-time application can be differentiated into two types: rigid and soft real-time. In rigid real time system, one-way delay tolerance is not allowed. The arrival of a packet after its specified end-to-end delay is considered as default of system. On other hand, in soft real time system, a probabilistic QOS guarantee is expected and delay is tolerated. Therefore, QoS for Real-time application of WSNs should guarantee deterministic or probabilistic end-to-end latency. For real-time application of WSNs, efficiency of energy cannot be neglected because the sensor nodes have a limited energy. Sensor nodes radio can be in active or sleep state to make it energy efficient, if nodes are in sleep state, and they are not able to receive or transmit data during that period. These active or sleep states do not favour the use of sensor networks in rigid real- time data delivery. The MAC layer provides channel access to the next hop while the network layer provides the end-to-end transmission time. Using cross-layer design approach optimum results can be obtained [15].

Motivation: The guaranteed delivery of data with low end-to-end delay and energy-efficiency is the most demanding requirement in industrial applications of WSNs. The traditional QoS routing protocol based on tree based routing are inefficient for dynamic network topology/asymmetric link characteristics. The geographically based routing achieve maximum packet progress towards destination within end-to-end delay. However, it considers only geographic distance rather than number of hops in the expenditure of energy and delay. Shortest-path-first maintains a list of optimal routes between the source and destination satisfying either reliability or timeliness. Hence, it is required to develop a real-time QoS routing protocol that guarantees packet delivery within a deadline and lower energy usage resulting in enhanced lifetime of the network.

Contribution: We have proposed RRDVCR protocol and the contributions are listed as follows: (i) Optimum routing path between source and the destination node is achieved in terms of the number of hops using virtual coordinates routing. (ii) Introduce a new dynamic weighting factor that analyses packets differently, depending on the remaining time to meet the end-to-end delay. (iii) A QoS parameter Maximum Number of Transmission Count [MTX] indicates link quality between the two nodes. (iv) Simulation performance comparison demonstrates low control overhead and energy consumption of the proposed RRDVCR protocol. Organization: The structure of a paper as follows: Section 2 reviews on related work. Section 3 gives system model, mathematical model and problem definition. The proposed protocol is explained in section 4. Simulation parameters are listed in section 5. The simulation results discussed in section 6. Conclusions are contained in section 7.

## II. Related work

This section summarizes the state of the research work emphasizing the QoS-aware routing protocols. Jalal et al.,[9] propose Multipath Routing Protocol which is Energy-efficient and QoS Aware, it maximizes lifetime by balancing the energy utilization across several nodes; it uses differentiation of service to grant packet that are sensitive to reach destination within the specified delay. However, multipath routing is not suitable because sending packets over multiple path inevitably incurs significant energy cost.

Li et al.,[2] use Two-hop neighbourhood information to select the next forwarder node. The packets are routed based on Two- hop velocity and energy utilization. This algorithm reduces packet deadline miss ratio. However, this work enhances lifetime of network at the cost of more energy utilization. Prabh et al.,[3] designed Transmission Scheduling Algorithm for Hexagonal Networks. That ensures that bottleneck node does not idle, with implicit clock synchronization to facilitate scheduling. However, the proposed scheduling algorithm works for only certain topology. Chungetal.,[21] use k-hop neighbor information for geographic packet routing with promising enhancement in routing delay and energy consumption in transmissions. However, as with increase in number of hops, it have low packet reception rate (PRR) and high retransmission cost. Marco et al.,[5] propose Three-hop Horizon Pruning(THP) to minimize the collision due to broadcasting. The algorithm computes a Connected Dominating Set (TCDS) where each node selects the subset of its one-hop neighbours which in-turn use its two-hop neighbours to reach next nodes which are three hops away. However, it is not feasible to determine link quality between nodes. The drawback is that, frequent HELLO messages have to be exchanged to measure the link quality. Wei et al.,[7] addressed the problem of finding optimum number of forward nodes, with complete dominant and partial dominant pruning. It uses two-hop neighbourhood information to minimize redundant transmissions. However, a intermediate node determine whether to rebroadcast the packet based on termination criterion that guarantees delivery. If no such termination criteria exists, then all packets are dropped. Katia et al., [6] propose Medium Access protocol with Traffic-Adaptive (TRAMA) a collision-free channel access is provided in Wireless Sensor Networks. Based on traffic, it uses election scheme at each node to find the time slot. Each node switches a low-power, idle state. However, proposed protocol is suited for delivery guarantees and energy efficiency than delay sensitive applications. The protocols [8][11][12] achieve real-time end-to-end delay requirement by selecting the next forwarding node based on the velocity offered by one-hop neighbourhood. Joseph et al.,[4] propose Low-media Access scheme to minimize duty cycle and idle time. However, it does not address packet delivery guarantee and reliable data transmission. Bile et al.,[25] designed ZigBee routing protocol that uses Hierarchical Tree Routing [HTR]. It meets the end-to-end delay requirements but it consumes more energy. The real-time QoS routing protocol for WSNs designed in [17]. It determine least cost paths using extended Dijkstra's shortest path algorithm for requirements of both real-time and non-real-time during connection establishment. Different paths are chosen for traffic of real-time and non-real-time. It uses a queuing model to serve both traffic. However, it does not consider asymmetric nature of channel, and the priority based queuing is too complex for resource constrained sensor nodes. A modified version of AODV protocol is proposed in [16], with prioritized packets based on its urgency. Packets with rigid end-to-end delay requirement are assigned with higher priority and are allowed through critical energy nodes and improves the network lifetime. However, the protocol has higher packet loss ratio. Geographic Opportunistic Routing [18] selects and prioritizes the forwarding node that meets required delay and end-to-end reliability. It uses one-hop neighbourhood information for determining node position. There is an increase in the control packets resulting in higher energy consumption. Gradient Routing with Two-hop [19] uses selective acknowledgement scheme to update the neighbour information. How- ever, the protocol does not uses link reliability between the nodes while making routing decision. Jianwei et al.,[20] propose Reliable Reactive Routing Enhancement [R3E] protocol that use biased backup scheme to determine guide path. It optimizes the packet delivery ratio and minimize the energy utilization. However, construction and maintenance of virtual path during route discovery introduces high overhead.

III. SYSTEM MODEL AND PROBLEM STATEMENT

A. *Virtual coordinate Routing for WSN*

The proposed algorithm adopts virtual coordinate routing protocol which enhances packet delivery reliability. The sink node builds reverse path tree by propagating advertisement (ADV) packets to gather the data from the sensor nodes, the ADV packet contain variables: sinkID, sourceID, residual energy, link quality(MTX) value and height-count. In each ADV packet, the sink node is set to height-count to 1. The height-count at the node is the minimum energy oriented and minimum number of hops to forward the packet from the $i^{th}$ node to the sink. Upon obtaining the ADV packet, each node's height is set which is equal to height-count in the packet and increments height-count by 1 and rebroadcast the ADV packets to its neighbour. The height-count of each node is indicated as Nh. The height-count for the sink node is set to 0. The height-count of source node and any node *i* is indicated as $h_s$ and $h_i$.

B. *Reliability Estimation Model*

Several methods are introduced to indicate link quality they are: Strength of Received Signal (RSS), Quality Indicator of Link (LQI) and Expected Transmission Count (ETX). The measurement of RSS, LQI are not accurate because of noise and interfering transmission [23]. Hence, they are not considered as link-quality metric. RSS is a signal-based indicator, and it is not relevant to the received packets. Therefore, RSS cannot be used as metric to express the link-quality. The LQI is another metric to indicate the link-quality, the LQI is a built- in parameter in CC2420 [7] chip that is used in most wireless sensor node, LQI uses the average correlation value of RSS for each receiving packets. Expected Transmission Count (ETX) is the anticipated value of transmissions that transmit packets successfully over wireless links in a both direction.. The forward delivery ratio $d_t$ is represent the probability of packets received successfully at the receiver. The reverse delivery ratio $d_r$ represent the acknowledgement packets reception of at the sender. The probability of successful delivery of a packet at receiver and its acknowledgement at sender is given as:

$$P_{succ} = d_r \times d_t \quad (1)$$

The ETX is probability of successful transmissions [23], as shown below

$$ETX = \frac{1}{P_{succ}} = \frac{1}{d_r \times d_t} \quad (2)$$

Although ETX metric is very efficient, However this metric is based on the behaviour of the link (E[Psuc]), but ETX does not check whether the current success probability (curPsuc) results in better delivery of packet. Whenever current success probability (curPsuc) is less than the maximum number of re-transmission permitted by MAC-layer, probability that delivered packets are discarded because, number of re-transmissions is increased. Therefore, it is necessary to consider account of the maximum number of re-transmissions permitted by the MAC layer. The physical layer properties and important in measuring the link quality in wireless networks to improve the performance. This paper defines the new metric called Maximum Transmission Count [MTX] which denotes the required number of transmissions on a link by considering the maximum number of re-transmissions permitted by the MAC-layer (MaxRetry). Minimum success probability Pl(i,j) offered by link between nodes $i$ and $j$) is denoted as

$$Pl(i,j) = \frac{1}{MaxRetry} \quad (3)$$

If current success probability curPsuc(i,j) is less than the minimum success probability Pl(i,j) offered by link between nodes i and j, the MTX metric is then defined as

$$MTX(i,j) = \begin{cases} \frac{1}{cur\,P_{succ}(i,j)} & \text{for } cur\,P_{succ}(i,j) \geq Pl(i,j) \\ \frac{1}{Pl(i,j)} & \text{for } cur\,P_{succ}(i,j) \leq Pl(i,j) \end{cases} \quad (4)$$

C. *Forwading Metric*

For node $i$, $N(i)$ is used to denote one hop neighbours node set. The source and sink nodes are labelled by *S, Dest* respectively. *N1 (i)* consists of the one-hop neighbours in one-hop area. *N2 (i)* is two-hop neighbours of node $i$. The number of hops between nodes $i$ and $j$ is represented by *Nh(i,j)*. Therefore, let *Sp* represent the expected packet progress towards sink for a given end-to-end delay *Dreq*.

$$S_p = \frac{N_h(S, Dest)}{D_{req}} \quad (5)$$

Where *Nh(S, Dest)* represents the Number of hops from the Source to the Sink.

$FN^1_{(i \to j)}(i)$ denotes one-hop forwarding node (FN) among the neighbours of node $i$ and can forward packets towards the sink. Let $C(1,i)$ denotes such available one-hop forwarder set for node $i$.

$$FN^1_{(i \to j)}(i) = \{j \in N_1 : N_h(i, Dest) - N_h(j, Dest)\} \quad (6)$$

Let $FN^2_{(i,j \to k)}(i)$ denotes two-hop forwarding nodes for node i, and can forward packets towards the sink. Let $C(2,i)$ denotes such available two-hop forwarder set for node i.

$$FN^2_{(i,j \to k)}(i) = \{j \in FN^1_{(i \to j)}(i) : N_h(i, Dest) - N_h(k, Dest)\} \quad (7)$$

The required packet progress speed offered by two-hop neighbors is determined. Thus $Sp(i,j,k)$ be the set of two-hop neighbors that provides required two-hop packet speed, which is calculated as:

$$S_p((i,j) \to k) = \frac{N_h(i, Dest) - N_h(k, Dest)}{ed_{(i,j)} + ed_{(j,k)}} \quad (8)$$

where ed(i,j) is the expected channel/medium latency from node $i$ to node $j$ and ed(j,k) is expected media latency from node $j$ to node $k$. Any (k) two-hop neighbour node of node i that has packet progress greater than $S_p((i,j) \to k)$ is selected as a forwarder and it is included in a set called potential Forwarder Node (FN$_{set}$) By taking residual energy level and the speed offered by forwarding nodes, THVR [13] has the following definition:

$$ve_{(i,j,k)} = \beta \times \frac{S_p((i,j) \to k)}{\sum_{k \in (FN_{set})} S_p((i,j) \to k)} + (1 - \beta) \times \frac{E_j/E_j^0}{\sum(E_j/E_j^0)} \quad (9)$$

The second term indicates residual energy of the nodes and initial energy at the instant of forwarding node *(k)* and β is a weighting factor. End-to-End delay performance is achieved for greater value β. Otherwise, it distributes traffic to nodes that has higher energy level. However, by fixing the coefficient of β value, real time and non-real time packets are serviced without any priority. In real-time application, each packet has different priority and different remaining time to meet the end- to-end delay. Hence, it is required to have adaptive coefficient depending on the remaining time to satisfy the required end-to-end delay and prioritize the packets. We proposes a adaptive coefficient value.

$$sre(i,j,k) = f(rt) \times \frac{S_p((i,j) \to k)}{\sum_{k \in (FN_{set})} S_p((i,j) \to k)} + f(rt) \times \frac{MTX(i,j)}{\sum_{k \in (FN_{set})} MTX(j,k)} + (1 - f(rt)) \times \frac{E_k/E_k^0}{\sum(E_k/E_k^0)} \quad (10)$$

where rt = $D_{req}$-$t_j$ is the remaining time to satisfy the end-to-end delay of the packet, and $t_j$ is time required for a node j to forward the packet. The function *f(rt)* should satisfy following two condition,

  i. The value of *f(rt)* ∈ *[0,1]* and
  ii. *f(rt)* is an inverse function. Based on above conditions, the f(rt) function can be

$$f(rt) = \begin{cases} \dfrac{rt}{D_{req}} & \text{for } \dfrac{rt}{D_{req}} \leq \dfrac{h_i}{h_s} \\ 1 - \dfrac{rt}{D_{req}} & \text{for } \dfrac{rt}{D_{req}} \geq \dfrac{h_i}{h_s} \end{cases} \quad (11)$$

Whenever the node wants to send a data packet, it identifies the forwarding node based on (13). In this work, when a sensor node has less remaining time to meet the end-to-end delay requirement of a packet, it finds the forwarder in its two-hop neighborhood that has higher speed of packet progress and it is more reliable. If sensor node has more remaining time to meet delay requirement, then sensor node selects forwarder node among its two-hop neighbors that has higher residual energy. If there are no nodes in two-hop neighbors that satisfy the required speed and more reliable then the node position is checked to find out the node that is near to sink and has higher residual energy and, it is worthwhile to forward the packet. If there is no node in two-hop neighborhood of sender that does not satisfy the required conditions, then the packets are dropped. In other scenario, the success probability of packet (MTX values), speed of packet progress and current status of energy availability at node is updated for every 2s. This update information is used to select the forwarding nodes among the two-hop neighbors of node. These process is repeated for 7 times. If the total number of attempt to transmit the packet is exceed 7, then packet is dropped. The maximum number of re-transmissions of a packet in each node is based on the link loss rate.

### D. Problem Statement

We formulate real time reliable data delivery using virtual coordinating routing for Wireless Sensor Networks (WSNs) as diversified objective with various constraints optimization problem

$$\text{Find } \left\{ \prod_k \left( \max(sre(i,j,k)) \right) \right\}$$
$$\text{Where } k \in FN^2_{(i,j \to k)}(i) \subseteq C_{(2,i)}$$

Subject to

$$ed\left(FN^2_{(i,j \to k)}(i)\right) \leq D_{req}$$
$$MTX(i,j) \geq Pl(i,j) \quad E_{(k,res)} \geq E_{th} \text{ where } k \in FN^2_{(i,j \to k)}(i)$$

The objective of the paper is to determine the optimal number of potential two-hop Forwarding Nodes(FNs) that is more re-liable, offers required packet advancement towards destination and has maximum residual energy.

### E. Media-Delay Estimation model

Let $T_k$ be the single hop channel delay of $j^{th}$ candidate, the channel delay can includes the delay of back-off and transmission of data packet at the sender. The next part is the coordination delay of candidate that is the time required for the $k^{th}$ candidate to send acknowledge to sender. The one-hop channel delay is represented in Equation 12, and delay of the signal propagation is ignored

$$ed_{(i,j)} = Ti_{Boff} + Ti_{Data} + j(T_{SIFS} + T_{Ack}) \quad (12)$$

Where $Ti_{Boff}$ is random back-off time for the sender to capture the channel,

## IV. PROPOSED ALGORITHM

In the proposed algorithm, there are three modules: a two-hop neighbor nodes module to find two-hop neighbor nodes, a Maximum Transmission Count (MTX) module to determine link reliability, and a forwarding node metric x with dynamic co-relation factor f (RT) module to determine forwarding node. Initially, the hop count for the sink is set to zero. The number of hops between a pair of nodes i and j is represented by Nh(i,j). For a given node *i* its neighbor of one-hop nodes and two-hop nodes are determined using Function 1. In function 1, a sensor node sends HELLO packets to its neighbor nodes, in turn the neighbor nodes send HELLO packets to its neighbors. Among these one-hop neighbors of node *i*, Function 1 determines the subset of one-hop which cover maximum number of two-hop of node *i*. These subset of one-hop nodes are used to forward the packets to two-hop nodes. Whenever two or more one-hop neighbors covers the same number of two-hop neighbors, then selection of one-hop neighbor is on basis of residual energy and minimum end-to-end delay. When node i has packet to transmit, it finds set of potential forwarding nodes among the two-hop nodes. The packet progress towards destination offered by each two-hop neighbor node of i is determined based on equation 10 and its neighbor nodes, in turn the neighbor nodes send HELLO packets to its neighbors.

---

**Data:** i, $N_1$,i
**Result:** $C_2$,i
**Initialization:** $C_2$,i=0
**for** (i=1; i≤| $N_1$, i |;i++) **do**

    *Add i ∈ $N_1$(i) which covers max(uncovered ($N_2$ (i))) to set $C_2$,i*

    *Add i ∈ $N_1$(i) if uncovered ($N_2$ (i)) covered by i only*
    **if**($C_2$,i = = $C_2$,j ) **then**

        *Find Comb(max($E_{n, res}$ and min(ed) )*
        *Add(Comb(max($E_{n, res}$ and min(ed)) to set $C_{2, i}$*
    **end**
**end**

Function 1: Subset of two-hop nodes as a forwarding nodes

**Algorithm: RRDCVR Algorithm**

**Data:** N(i), $N_2(i)$, $N_h$(i, Dest), $N_h$(k, Dest), Dest, i
**Result:** $FN^2_{(i,j->k)}(i)$, $FN^1_{(i->j)}(i)$
Initialization: $FN^2_{(i,j->k)}(i) = 0$, $E_{avl} = 5J$ $h_{Dest} = 0$

$$D_{req} = \frac{N_h(i, Dest)}{ed(i, Dest)}$$

**for each** k ∈ $FN^2_{(i,j->k)}(i)$ **do**

$$S_p((i,j) \to k) = \frac{N_h(i,Dest) - N_h(k,Dest)}{ed(i,j) + ed(j,k)}$$

**if** $S_p((i,j) \to k) \geq D_{req}$ **then**
  $FN_{set} = S_p((i,j) \to k)$
  **for each** k ∈ $FN_{set}$ **do**
    **if**(MTX(MTX(i,j)≥ 0.5),k)≥ 0.5) **then**
      *Run Equation 10*
    **end**
    $return(\pi(Max(sre(i,j,k))))$
  **end**
**end**
*return*

**Function 2: MTX(i,j)**

**Data:** *MaxRetry, Psuc(x, y), $d_r$, $d_t$, minPsuc*
**Result:** MTX(i,j, y)
**Initialization:** MaxRetry = 7
*Psuc(x, y) = $d_r*d_t$*
**if** ( curPsuc(i,j) ≥ Pl(i,j)) **then**

$$MTX(i,j) = \frac{1}{curPsucc(i,j)}$$

**else**

$$MTX(i,j) = \frac{1}{P(i,j)}$$

**end**

## V. SIMULATION PARAMETERS

The proposed protocol RRDVCR is simulated and evaluated using ns-2 [25]. It is compared with THVR[13] and SPEED[8]. The simulation network consists of 200 nodes deployed in a 200mX200m area. Distribution of nodes follows Poisson point process with a density of 0.005 node / m2. The location of source nodes is at the region (15m, 25m) while the sink in the area (155m, 125m). The source generates a CBR flow of 1 packet/second and a packet size of 150 bytes. The energy consumption at MAC layer, link quality and parameters are set as per Mica2 Motes [24] with MPR400 radio as per THVR [13]. THVR [13] and SPEED [8] are QoS based protocols and main simulation parameters are PDMR (Packet Deadline Miss Ratio), ECPP (Energy Consumed Per Packet is the total energy utilized by the all packets transmitted successfully), the packet average delay (mean of packet delay) and worst case delay (largest value sustained by the successful transmitted packet) are obtained.

## VI. PERFORMANCE ANALYSIS

This section, the performance of the proposed algorithm RRDVCR based on simulation results are discussed. Fig. 1 shows the number of packets missed with given different end-to-end delay. In the beginning, miss ratio is between 78% and 80% in the proposed protocol while in THVR [13] protocol it is 80%. The packet loss increases as the forwarding nodes are not offering the required packet progress towards destinations. As end-to-end delay is increased, the forwarding nodes meet the expected end-to-end delay. With 1200s as end-to-end delay, it is noticed that number of packets missed is lower compared to the existing protocol as higher number of forwarding nodes are available to forward the packets. At interval 1500s to 1800s end-to-end delay, almost all the packets reach the destination. The number of packets missed is almost zero as void space between the forwarding nodes is less and more forwarding nodes are available with required speed.

Fig. 2 shows consumption of energy by each packet that are transmitted successfully from the originator node to the destintion node. In THVR [13], more number of control packets are exchanged since it uses proactive approach for updating two-hop neighbor information. The RRDVCR forward packets in smaller number of hops with reduced number of control packets and hence it requires less energy. It is observed from the figure that the consumption of energy is more during 900s to 1100s deadline.

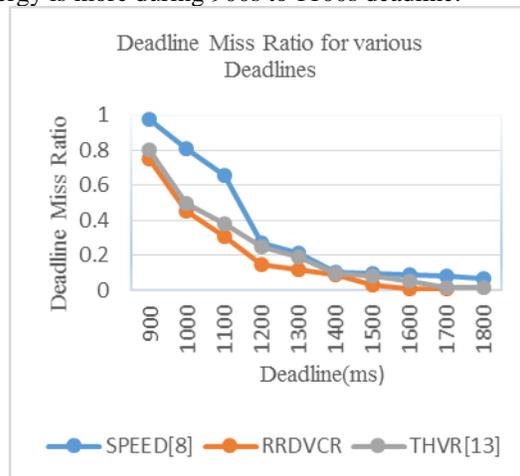

Fig. 1. Delivery Miss Ratio with varied End-to-End Delay. (i) SPEED; (ii) THVR; (iii) RRDVCR

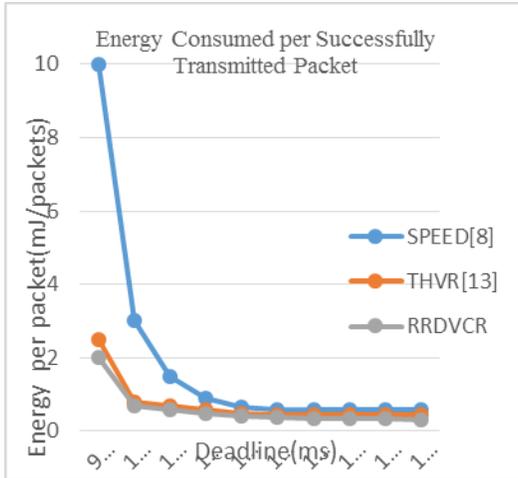

Fig. 2. Energy consumed per successfully transmitted packet. (i) SPEED; (ii) THVR; and (iii) RRDVCR

Fig. 3, shows the performance of RRDVCR with multiple source nodes. The probability of delivering packets successfully decreases as transmission from multiple source nodes generate huge traffic in networks that cause congestion, increase channel contention in the network resulting in higher packet collision and re-transmission of packets. It is observed that as the count of source nodes increase, the total number of packets unable to reach the destination also increases. The packet deadline miss ratio is about 23% in RRDVCR when number of sources in network are 4. As number of source nodes are increased from 4 to 6, the packet deadline miss ratio increase from from 23% to 24% in RRDVCR, whereas in THVR [13] it increases from 24% to 25%. In SPEED [8] protocol, the packet deadline miss ratio is about 50%, since it considers one-hop information to select forwarding nodes and hence packets travel more distance to reach destination.

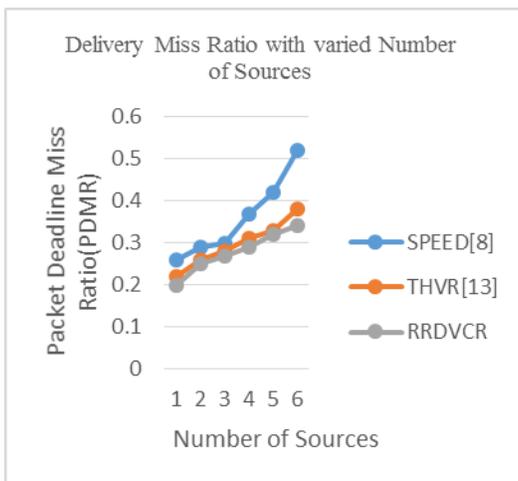

Fig. 3. Delivery Miss Ration with varied number of sources. for (i) SPEED; (ii) THVR; and (iii) RRDVCR; while the number of source nodes increases from 1 to 6.

Fig. 4 indicates the required energy for every packet transmission with varying number of source nodes on the networks. One common characteristic observed is that for all existing protocols and the proposed protocol, energy consumption during transmission of packets increases with the increase in the total number source nodes in network. The RRDVCR protocol is energy-efficient compared to existing protocols as it delivers packets in smaller number of hops with minimum number of control packets. RRDVCR uses piggyback concept to update link status between nodes. In THVR [13], energy consumption is more because of (i) two-hop information is updated frequently with control packets which requires more energy. (ii) The increase in number of source nodes results in more number of transmission of packets from different sources, which leads to congestion at different layers. The energy consumption is 46mJ in RRDVCR, 48mJ in THVR [13] and 58mJ in SPEED [8]. When sources are increases to 6, consumption of energy increases. It is 80mJ in SPEED [8], 66 mJ in THVR [13] and 49mJ in our protocol RRDVCR. It indicates that RRDVCR results in lower value of deadline miss ratio and lower consumption of energy for each successfully transmitted packet. It is 74% more efficient than THVR[13]and SPEED[8].

Fig. 5 shows impact of co-relation factor f(rt) on network performance. When the co-relation factor f(rt)is set to 0.1, the co-relation factor favors energy balance and ignores the end-to-end delay. Packets unable to reach destination within the specified end-to-end delay are more. When the co-relation factor is set to 0.1, with a rigid end-to-end delay(900 ms), 80% of packets are unable to reach the destination. As end-to-end delay is relaxed, the forwarding nodes get time to forward more number of packets and the packet miss ratio gradually reduces. Whenever co-relation factor f(rt) is set to 0.9, it meets the end-to-end delay requirement rather than energy balance. Every packet has its own remaining time to satisfy the deadline and each node services a packet differently and hence the co-relation factor f(rt) is varied dynamically. The packet miss ratio decreases to 7% to 8%

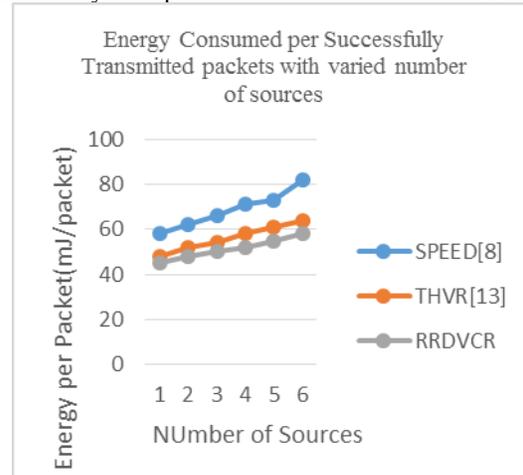

Fig. 4. Energy consumed per successfully transmitted packet with varied number of source nodes. (i) SPEED; (ii) THVR; and (iii) RRDVCR

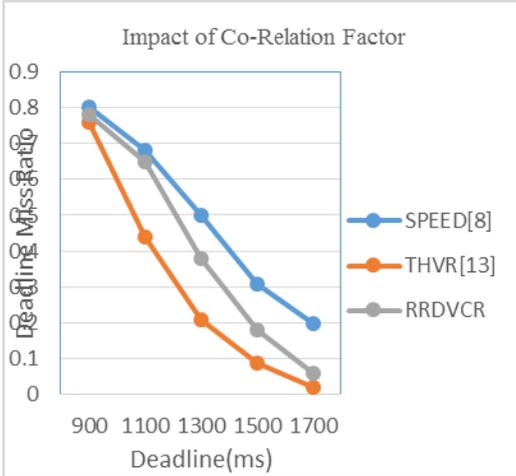
Fig. 5. Impact of different value of co-relation factor.

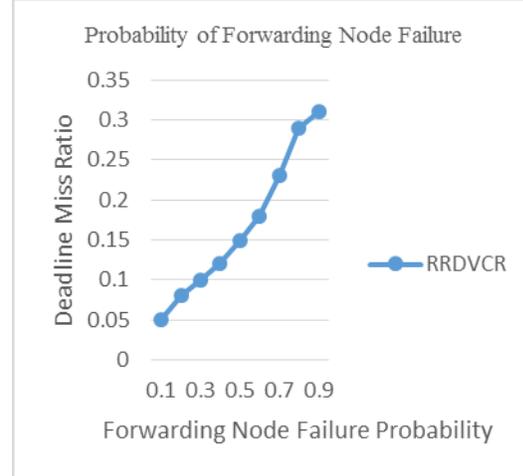
Fig. 6. Probability of node failure with Delivery Miss Ratio.

Fig. 6 shows number of packets miss rate before the forwarding nodes fail to forward packets. The maximum tolerance for probability of forwarding node failure is set to 0.5 in our simulation. Whenever the probability of forwarding node failure exceeds 0.5, then such forwarding node is called as dead node. The probability of forwarding node failure means either it does not satisfy the packet progression speed or the residual energy is less than the threshold energy.

Fig. 7 illustrates the number of hops that exist between the source and the destination. Optimum number of hops between source and destination is achieved in RRDVCR. When the network has 10 nodes, the hops between source to destination is about 4 which is 60% of the total network size. This is because network is sparse and nodes are placed at a far distance, to transfer data packets between the source and the destination; 60% of intermediate nodes act as router to forward packets. In THVR[13] the number of hops between source and destination is more when compared to proposed algorithm; this is because it does not estimate the success probability between nodes and the forwarding node selection is based only on geographic distance progress offered by two-hop neighbors. the number of hops between the source and the destination for network size 20,30,40 is 9,12,17 in proposed algorithm and in THVR[13] it is 13,19,26. This difference is due to the intermediate nodes having its neighbor node's probability of success, and neighbor node packet progress towards destination in-terms of number of hops. As simulation progresses all the nodes get update of success probability of its neighbor node and number of hops remaining to reach destination, the nodes in proposed algorithm converge fast as compared to THVR [13]. Because of these reasons, the number of hops between the source and the destination is less in the proposed algorithm.

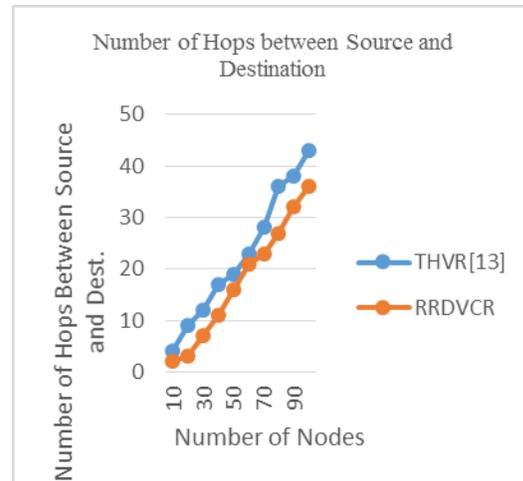
Fig. 7. Comparison of Number of Hopes. (i) RRDVCR; and (ii) THVR[13]

## VII. CONCLUSION

The Proposed RRDVCR is based on virtual coordinate routing. The proposed protocol employ a link quality estimation method (MTX:Maximum Transmission Count) which helps in selecting the link that has high reliability. The link reliability, packet advancement and residual energy of node are the parameters taken into account during selection potential forwarding nodes. The dynamic co-relation factor is introduced in this paper that co-relates different metric during potential forwarding node selection and the real- time or non real-time packets are serviced accordingly remaining time meet the deadline. The simulation results shows improvement in packet success delivery ratio within specified deadline, energy consumption during transmission of packet for varied number of source node is decreased. With dynamic co-relation factor, number of packet missed to reach destination within end-to-end delay is reduced.

| Your Name | Title* | Research Field | Personal website |
|---|---|---|---|
| Venkatesh | Asst.Prof | WSN | |
| CS Sengar | Student | WSN | |
| Venugopal K R | Professor | WSN | www.venugopalkr.com |